%% file: main.tex
\pdfoutput=1
\documentclass[12pt,a4paper]{article}

\usepackage{ifthen} 
\newboolean{pdflatex}
\setboolean{pdflatex}{true} 

\newboolean{articletitles}
\setboolean{articletitles}{true} 

\newboolean{uprightparticles}
\setboolean{uprightparticles}{false} 

\usepackage{booktabs}

\def\phih       {{\ensuremath{\phi}}\xspace}

\def\BdbToKstbMuMu {\decay{\Bzb}{\Kstarzb\mumu}}

\def\DTV {\delta\phi_{\mathcal{C}_9}}
\def\DTAV {\delta\phi_{\mathcal{C}_{10}}}

\DeclareRobustCommand{\optmybar}[1]{\shortstack{{\miniscule (\rule[.3ex]{0.65em}{.18mm})}
  \\ [-.7ex] $#1$}}
\def\bnbar#1  {\kern \thebaroffset\optmybar{\kern -\thebaroffset \ensuremath{#1}}\xspace}

\def\paperauthors{LHCb collaboration} 
\def\paperasciititle{CP violation analysis of local and nonlocal amplitudes in the in the B-bar to Kstar-bar mu mu decay} 
\def\papertitle{\CP violation analysis of local and nonlocal amplitudes in the $\Bzb \to \Kstarzb\mu^+\mu^-$ decay} 
\def\paperkeywords{{High Energy Physics}, {LHCb}, {Penguin decays}, {CP violation}, {Wilson coefficients}} 
\def\papercopyright{\the\year\ CERN for the benefit of the LHCb collaboration} 
\def\paperlicence{CC-BY-4.0 licence}
\def\paperlicenceurl{https://creativecommons.org/licenses/by/4.0/}

\input{preamble}

\begin{document}

\renewcommand{\thefootnote}{\fnsymbol{footnote}}
\setcounter{footnote}{1}

\input{title-LHCb-PAPER}


\renewcommand{\thefootnote}{\arabic{footnote}}
\setcounter{footnote}{0}

\cleardoublepage


\pagestyle{plain} 
\setcounter{page}{1}
\pagenumbering{arabic}


\input{body}

\input{acknowledgements}

\addcontentsline{toc}{section}{References}

\bibliographystyle{LHCb}
\bibliography{main,standard,LHCb-PAPER,LHCb-CONF,LHCb-DP,LHCb-TDR}

\newpage
\appendix

\newpage
\input{Authorship_LHCb-PAPER-2026-007}

\end{document}

%% file: preamble.tex
\usepackage[top=1in, bottom=1.25in, left=1in, right=1in]{geometry}

\usepackage{microtype}
\usepackage{lineno}  
\usepackage{xspace} 
\usepackage{caption} 

\usepackage{graphicx}  
\usepackage{color}
\usepackage{colortbl}
\graphicspath{{./figs/}} 

\usepackage{amsmath} 
\usepackage{amssymb}
\usepackage{amsfonts}
\usepackage{upgreek} 

\newcommand*\patchAmsMathEnvironmentForLineno[1]{%
\expandafter\let\csname old#1\expandafter\endcsname\csname #1\endcsname
\expandafter\let\csname oldend#1\expandafter\endcsname\csname
end#1\endcsname
 \renewenvironment{#1}%
   {\linenomath\csname old#1\endcsname}%
   {\csname oldend#1\endcsname\endlinenomath}%
}
\newcommand*\patchBothAmsMathEnvironmentsForLineno[1]{%
  \patchAmsMathEnvironmentForLineno{#1}%
  \patchAmsMathEnvironmentForLineno{#1*}%
}
\AtBeginDocument{%
\patchBothAmsMathEnvironmentsForLineno{equation}%
\patchBothAmsMathEnvironmentsForLineno{align}%
\patchBothAmsMathEnvironmentsForLineno{flalign}%
\patchBothAmsMathEnvironmentsForLineno{alignat}%
\patchBothAmsMathEnvironmentsForLineno{gather}%
\patchBothAmsMathEnvironmentsForLineno{multline}%
\patchBothAmsMathEnvironmentsForLineno{eqnarray}%
}

\usepackage[pdftex,
            pdfauthor={\paperauthors},
            pdftitle={\paperasciititle},
            pdfkeywords={\paperkeywords}]{hyperref}

\usepackage{hyperxmp}

\hypersetup{
  pdfcopyright={Copyright (C) \papercopyright},
  pdflicenseurl={\paperlicenceurl}
}

\usepackage[bottom,flushmargin,hang,multiple]{footmisc}

\usepackage[all]{hypcap} 

\input{lhcb-symbols-def} 

\usepackage{cite} 
\usepackage{mciteplus}


%% file: lhcb-symbols-def.tex
\usepackage{xspace} 
\usepackage{upgreek}

\def\lhcb   {\mbox{LHCb}\xspace}

\def\MagUp {\mbox{\em Mag\kern -0.05em Up}\xspace}

\ifthenelse{\boolean{uprightparticles}}%
{

 \def\Pmu         {\ensuremath{\upmu}\xspace}

 \def\Ppi         {\ensuremath{\uppi}\xspace}

 \def\Ppsi        {\ensuremath{\uppsi}\xspace}

 \def\PDelta      {\ensuremath{\Delta}\xspace}                 
 \def\PXi         {\ensuremath{\Xi}\xspace}                 
 \def\PLambda     {\ensuremath{\Lambda}\xspace}                 
 \def\PSigma      {\ensuremath{\Sigma}\xspace}                 
 \def\POmega      {\ensuremath{\Omega}\xspace}                 
 \def\PUpsilon    {\ensuremath{\Upsilon}\xspace}
 \let\oldPi\Pi
 \def\PPi         {\ensuremath{\oldPi}\xspace}

 \def\PB      {\ensuremath{\mathrm{B}}\xspace}                 
                  
 \def\PD      {\ensuremath{\mathrm{D}}\xspace}

 \def\PJ      {\ensuremath{\mathrm{J}}\xspace}                 
 \def\PK      {\ensuremath{\mathrm{K}}\xspace}

 \def\Pb      {\ensuremath{\mathrm{b}}\xspace}

 \def\Pi      {\ensuremath{\mathrm{i}}\xspace}

 \def\Pp      {\ensuremath{\mathrm{p}}\xspace}

 \def\Ps      {\ensuremath{\mathrm{s}}\xspace}

 \def\thebaroffset{0.0em}
}
{

 \def\Pmu         {\ensuremath{\mu}\xspace}

 \def\Ppi         {\ensuremath{\pi}\xspace}

 \def\Ppsi        {\ensuremath{\psi}\xspace}                 
                  
 \mathchardef\PDelta="7101
 \mathchardef\PXi="7104
 \mathchardef\PLambda="7103
 \mathchardef\PSigma="7106
 \mathchardef\POmega="710A
 \mathchardef\PUpsilon="7107
 \mathchardef\PPi="7105
                  
 \def\PB      {\ensuremath{B}\xspace}                 
                  
 \def\PD      {\ensuremath{D}\xspace}

 \def\PJ      {\ensuremath{J}\xspace}                 
 \def\PK      {\ensuremath{K}\xspace}

 \def\Pb      {\ensuremath{b}\xspace}

 \def\Pi      {\ensuremath{i}\xspace}

 \def\Pp      {\ensuremath{p}\xspace}

 \def\Ps      {\ensuremath{s}\xspace}

 \def\thebaroffset{0.18em}
}
\newcommand{\offsetoverline}[2][\thebaroffset]{\kern #1\overline{\kern -#1 #2}}%

\makeatletter
\ifcase \@ptsize \relax
  \newcommand{\miniscule}{\@setfontsize\miniscule{4}{5}}
\or
  \newcommand{\miniscule}{\@setfontsize\miniscule{5}{6}}
\or
  \newcommand{\miniscule}{\@setfontsize\miniscule{5}{6}}
\fi
\makeatother

\DeclareRobustCommand{\optbar}[1]{\shortstack{{\miniscule (\rule[.5ex]{1.25em}{.18mm})}
  \\ [-.7ex] $#1$}}

\def\mup        {{\ensuremath{\Pmu^+}}\xspace}
\def\mun        {{\ensuremath{\Pmu^-}}\xspace} 

\def\mumu       {{\ensuremath{\Pmu^+\Pmu^-}}\xspace}

\def\squark    {{\ensuremath{\Ps}}\xspace}

\def\bquark    {{\ensuremath{\Pb}}\xspace}

\def\pion   {{\ensuremath{\Ppi}}\xspace}

\def\pip    {{\ensuremath{\pion^+}}\xspace}
\def\pim    {{\ensuremath{\pion^-}}\xspace}

\def\kaon    {{\ensuremath{\PK}}\xspace}
\def\Kbar    {{\ensuremath{\offsetoverline{\PK}}}\xspace}

\def\KorKbar {\kern \thebaroffset\optbar{\kern -\thebaroffset \PK}{}\xspace}

\def\Kp      {{\ensuremath{\kaon^+}}\xspace}
\def\Km      {{\ensuremath{\kaon^-}}\xspace}

\def\Kstarz  {{\ensuremath{\kaon^{*0}}}\xspace}
\def\Kstarzb {{\ensuremath{\Kbar{}^{*0}}}\xspace}

\def\Kstarm  {{\ensuremath{\kaon^{*-}}}\xspace}

\def\Dbar    {{\ensuremath{\offsetoverline{\PD}}}\xspace}
\def\D       {{\ensuremath{\PD}}\xspace}

\def\DorDbar {\kern \thebaroffset\optbar{\kern -\thebaroffset \PD}\xspace}

\def\Dp      {{\ensuremath{\D^+}}\xspace}
\def\Dm      {{\ensuremath{\D^-}}\xspace}

\def\DpDm    {\ensuremath{\Dp {\kern -0.16em \Dm}}\xspace}

\def\B       {{\ensuremath{\PB}}\xspace}
\def\Bbar    {{\ensuremath{\offsetoverline{\PB}}}\xspace}

\def\BorBbar {\kern \thebaroffset\optbar{\kern -\thebaroffset \PB}\xspace}
\def\Bz      {{\ensuremath{\B^0}}\xspace}
\def\Bzb     {{\ensuremath{\Bbar{}^0}}\xspace}
\def\Bd      {{\ensuremath{\B^0}}\xspace}
\def\Bdb     {{\ensuremath{\Bbar{}^0}}\xspace}
\def\BdorBdbar {\kern \thebaroffset\optbar{\kern -\thebaroffset \Bd}\xspace}

\def\Bub     {{\ensuremath{\B^-}}\xspace}

\def\Bs      {{\ensuremath{\B^0_\squark}}\xspace}
\def\Bsb     {{\ensuremath{\Bbar{}^0_\squark}}\xspace}
\def\BsorBsbar {\kern \thebaroffset\optbar{\kern -\thebaroffset \Bs}\xspace}

\def\jpsi     {{\ensuremath{{\PJ\mskip -3mu/\mskip -2mu\Ppsi}}}\xspace}
\def\psitwos  {{\ensuremath{\Ppsi{(2S)}}}\xspace}

\def\Y#1S{\ensuremath{\PUpsilon{(#1S)}}\xspace}

\def\proton      {{\ensuremath{\Pp}}\xspace}

\def\Lz          {{\ensuremath{\PLambda}}\xspace}

\def\LorLbar     {\kern \thebaroffset\optbar{\kern -\thebaroffset \PLambda}\xspace}

\def\Lb           {{\ensuremath{\Lz^0_\bquark}}\xspace}

\newcommand{\decay}[2]{\ensuremath{#1\!\to #2}\xspace} 

\def\to                 {\ensuremath{\rightarrow}\xspace}

\def\qsq       {{\ensuremath{q^2}}\xspace}

\def\CP                {{\ensuremath{C\!P}}\xspace}

\def\bsll     {\decay{\bquark}{\squark \ell^+ \ell^-}}

\def\AT#1     {\ensuremath{A_{\mathrm{T}}^{#1}}\xspace}           

\def\ctl       {\ensuremath{\cos{\theta_\ell}}\xspace}
\def\ctk       {\ensuremath{\cos{\theta_K}}\xspace}

\def\C#1      {\ensuremath{\mathcal{C}_{#1}}\xspace}                       
\def\Cp#1     {\ensuremath{\mathcal{C}_{#1}^{'}}\xspace}                    
\def\Ceff#1   {\ensuremath{\mathcal{C}_{#1}^{\mathrm{(eff)}}}\xspace}        
\def\Cpeff#1  {\ensuremath{\mathcal{C}_{#1}^{'\mathrm{(eff)}}}\xspace}       
\def\Ope#1    {\ensuremath{\mathcal{O}_{#1}}\xspace}                       
\def\Opep#1   {\ensuremath{\mathcal{O}_{#1}^{'}}\xspace}                    

\newcommand{\aunit}[1]{\ensuremath{\text{\,#1}}}       

\newcommand{\tev}{\aunit{Te\kern -0.1em V}\xspace}
\newcommand{\gev}{\aunit{Ge\kern -0.1em V}\xspace}
\newcommand{\mev}{\aunit{Me\kern -0.1em V}\xspace}
\newcommand{\kev}{\aunit{ke\kern -0.1em V}\xspace}
\newcommand{\ev}{\aunit{e\kern -0.1em V}\xspace}
 
\newcommand{\mevc}{\ensuremath{\aunit{Me\kern -0.1em V\!/}c}\xspace}
\newcommand{\gevc}{\ensuremath{\aunit{Ge\kern -0.1em V\!/}c}\xspace}
\newcommand{\mevcc}{\ensuremath{\aunit{Me\kern -0.1em V\!/}c^2}\xspace}
\newcommand{\gevcc}{\ensuremath{\aunit{Ge\kern -0.1em V\!/}c^2}\xspace}
\newcommand{\gevgevcccc}{\ensuremath{\gev^2\!/c^4}\xspace} 

\def\fb   {\ensuremath{\aunit{fb}}\xspace}
\def\invfb   {\ensuremath{\fb^{-1}}\xspace}

\newcommand{\stat}{\aunit{(stat)}\xspace}
\newcommand{\syst}{\aunit{(syst)}\xspace}

\def\deriv {\ensuremath{\mathrm{d}}}

\def\gsim{{~\raise.15em\hbox{$>$}\kern-.85em
          \lower.35em\hbox{$\sim$}~}\xspace}
\def\lsim{{~\raise.15em\hbox{$<$}\kern-.85em
          \lower.35em\hbox{$\sim$}~}\xspace}

\newcommand{\abs}[1]{\ensuremath{\left\|#1\right\|}} 
\newcommand{\Real}{\ensuremath{\mathcal{R}e}\xspace}
\newcommand{\Imag}{\ensuremath{\mathcal{I}m}\xspace}

\def\evtgen     {\mbox{\textsc{EvtGen}}\xspace}

\def\geant      {\mbox{\textsc{Geant4}}\xspace}

\def\photos     {\mbox{\textsc{Photos}}\xspace}

\def\pythia     {\mbox{\textsc{Pythia}}\xspace}

\def\tell1  {TELL1\xspace}
\def\ukl1   {UKL1\xspace}

\newcommand{\ie}{\mbox{\itshape i.e.}\xspace}

\newcommand{\lhcborcid}[1]{\href{https://orcid.org/#1}{\hspace*{0.1em}\raisebox{-0.45ex}{\includegraphics[width=1em]{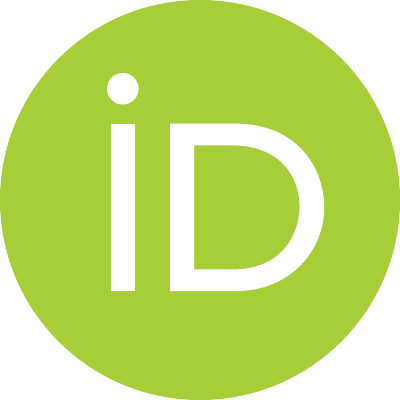}}}}


%% file: title-LHCb-PAPER.tex

\begin{titlepage}
\pagenumbering{roman}

\vspace*{-1.5cm}
\centerline{\large EUROPEAN ORGANIZATION FOR NUCLEAR RESEARCH (CERN)}
\vspace*{1.5cm}
\noindent
\begin{tabular*}{\linewidth}{lc@{\extracolsep{\fill}}r@{\extracolsep{0pt}}}
\ifthenelse{\boolean{pdflatex}}
{\vspace*{-1.5cm}\mbox{\!\!\!\includegraphics[width=.14\textwidth]{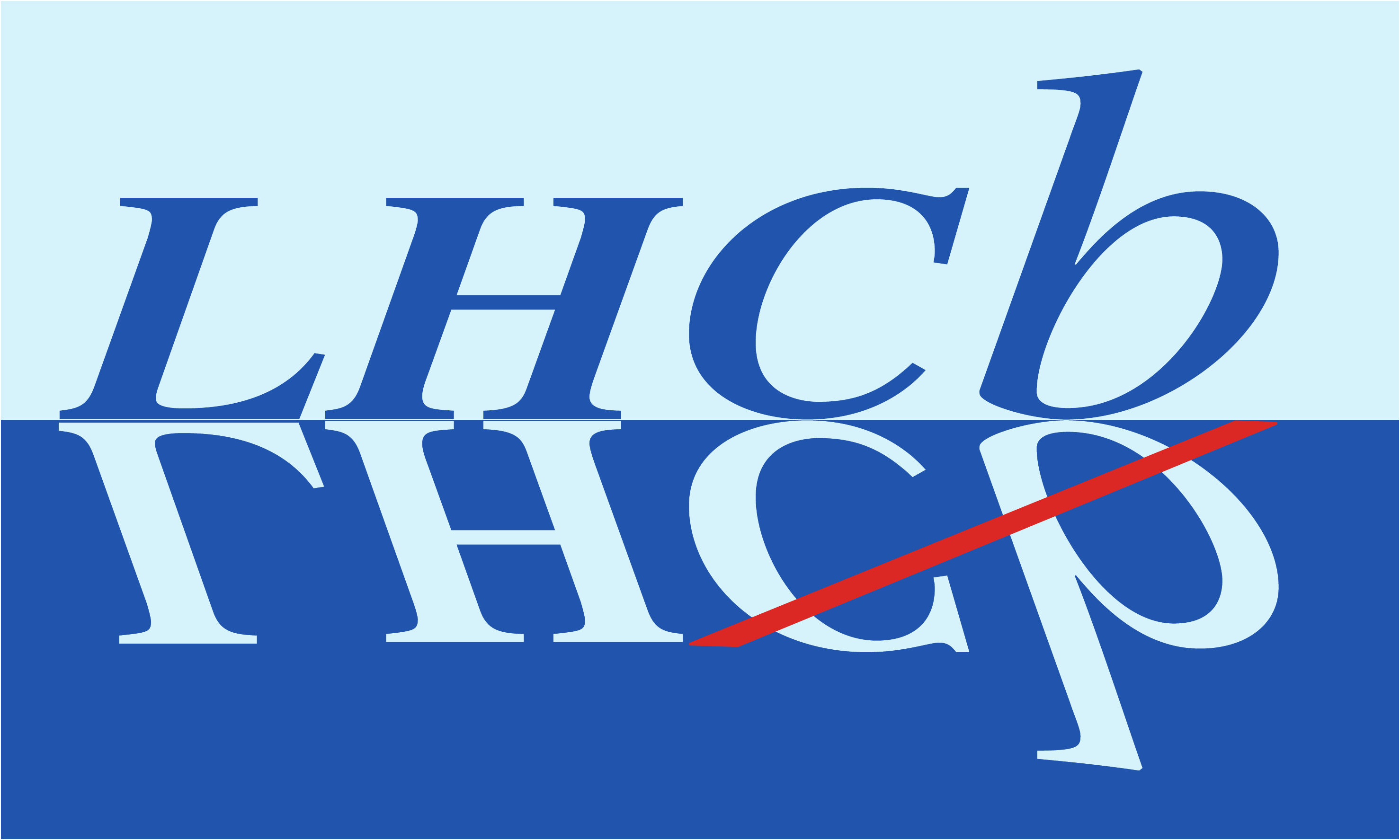}} & &}%
{\vspace*{-1.2cm}\mbox{\!\!\!\includegraphics[width=.12\textwidth]{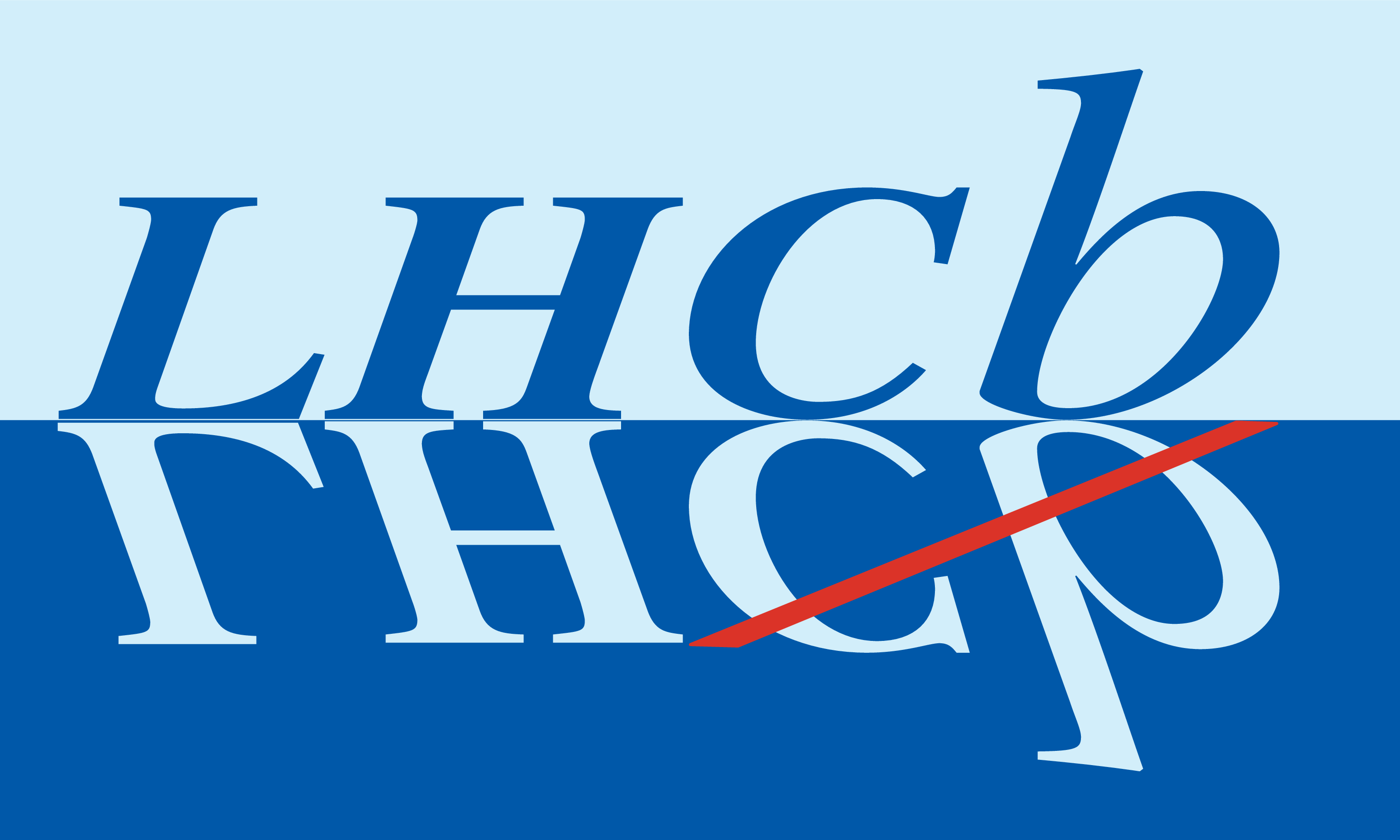}} & &}%
\\
 & & CERN-EP-2026-114 \\  
 & & LHCb-PAPER-2026-007 \\  
 & & August 10, 2026 \\
\end{tabular*}

\vspace*{2.0cm}

{\normalfont\bfseries\boldmath\huge
\begin{center}
  \papertitle 
\end{center}
}

\vspace*{1.0cm}

\begin{center}
\paperauthors\footnote{Authors are listed at the end of this Letter.}
\end{center}

\vspace{\fill}

\begin{abstract}
  \noindent
A search for \CP violation in the $\decay{\Bdb}{\Kstarzb\mup\mun}$ decay is performed using proton-proton collision data collected by the LHCb experiment during \mbox{Run 1} and \mbox{Run 2}, corresponding to an integrated luminosity of 8.4\invfb. The analysis exploits the full angular distribution of the decay, providing sensitivity to \CP-violating effects in both vector and axial-vector contributions to this flavour-changing neutral-current process. The complex Wilson coefficients are determined within the Weak Effective Theory through an unbinned maximum-likelihood fit to the angular observables, incorporating nonlocal hadronic amplitudes across the full dimuon mass spectrum. The precision of the \CP-violation observables is improved by an order of magnitude relative to previous measurements, with the imaginary parts of the Wilson coefficients now determined more precisely than the real parts. No significant \CP violation is observed, and the results are consistent with Standard Model.
\end{abstract}


\begin{center}
  Published in Phys. Rev. Lett. 137, (2026) 061801
\end{center}

\vspace{\fill}

{\footnotesize 
\centerline{\copyright~\papercopyright. \href{\paperlicenceurl}{\paperlicence}.}}
\vspace*{2mm}

\end{titlepage}


\newpage
\setcounter{page}{2}
\mbox{~}
%
%
%
%

%% file: body.tex
\noindent The study of differences between matter and antimatter is central to understand the observed dominance of matter in the Universe. These differences, potentially caused by \CP violation, can be probed by comparing a process to that obtained by applying the simultaneous Charge conjugation ($C$) and Parity ($P$) symmetry operations. Despite the remarkable success of the Standard Model~(SM) in describing particle interactions, its built-in mechanism for \CP violation is too weak to account for the observed asymmetry in the Universe, motivating searches for \CP-violating physics beyond the SM~(BSM)~\cite{Gavela:1993ts}.

Flavour-changing neutral-current~(FCNC) decays such as \BdbToKstbMuMu are very sensitive to BSM effects because of the suppressed nature of the SM contributions. Indeed, analyses of the decay over the past decade have produced results that are difficult to explain within the SM \cite{LHCb-PAPER-2025-041,CMS:2024atz}. However, due to the challenges in calculating the effects of the strong force, it is not possible to draw firm conclusions on the evidence for BSM. In Ref.~\cite{LHCb-PAPER-2024-011} we reported a comprehensive \CP-averaged amplitude analysis of \BdbToKstbMuMu over the full dimuon mass spectrum, including the charmonium regions. The analysis used a dispersive description of the one- and two-particle nonlocal amplitudes to model the long-distance contributions and their interference with the short-distance terms, enabling direct measurements of the Wilson coefficients. 
In this Letter, we extend this previous measurement analysing the dataset collected by the \lhcb experiment between 2011 and 2018 to include effects of \CP violation. An observation of \CP violation in this decay that significantly deviates from the SM expectation would lend itself to stronger conclusions about the presence of BSM physics and could form part of the explanation of the matter-antimatter asymmetry of the Universe. Throughout this Letter, charge-conjugate processes are implied unless explicitly stated otherwise.

The \BdbToKstbMuMu decay with \decay{\Kstarzb}{\Km\pip} is a flavour-specific final state, \ie the charge of the kaon uniquely identifies the flavour of the neutral \B meson at the time of decay. In this case, \CP violation can manifest itself through interfering amplitudes that in turn lead to different angular distributions and rates for the \Bdb and the corresponding \Bd decays. At any point in the phase space of the decay products, each amplitude has a strong phase, common to  the \Bdb and the \Bd decay, and a weak phase, responsible for the \CP violation, which changes sign between the two. A previous measurement based on data corresponding to 3\invfb from the \lhcb experiment~\cite{LHCb-PAPER-2014-032} analysed regions of dimuon mass squared ($\qsq$) far away from the \jpsi and \psitwos charmonium resonances that share the same final state. In those regions, the interfering amplitudes have both to be from the FCNC process, which all have the same strong phase. The implication is that there is only sensitivity to so-called $T$-odd asymmetries~\cite{Bobeth:2008ij}. In this Letter, we present an analysis that is unbinned in $\qsq$ with no charmonium vetoes. This means that we are sensitive to the interference between 
the amplitudes describing the charmonium resonances and the FCNC process. As the charmonium resonances have a phase shift that depends on $\qsq$, the analysis gains sensitivity to the 
$T$-even asymmetries as well, which are proportional to the sine of the strong phase difference~\cite{Blake:2017fyh}.

In the SM, the weak phase difference between the $\decay{b}{c}$ tree-level amplitude of the charmonium resonances and the $\decay{b}{s}$ FCNC amplitude arises from the complex phases of the elements of the Cabibbo--Kobayashi--Maskawa~(CKM) matrix and is given by $\arg(\lambda_t/\lambda_c)$, where $\lambda_q=V_{qb}V^*_{qs}$ for the quark $q$. This is the same SM phase that can be measured through time-dependent \CP violation in $\Bs$ decays to \CP eigenstates. However, if \CP-violating BSM effects are involved in either $\Bs$ oscillations or in the $\decay{b}{s\mumu}$ FCNC transitions, it is not expected that they would show up as the same weak phase in the two types of measurement.

The amplitudes describing the decay can be divided up into local and nonlocal components. The local ones are due to FCNC loops involving virtual particles with masses well above the mass of the \Bdb while the nonlocal amplitudes involve resonance decays such as \decay{\Bdb}{\jpsi\Kstarzb} followed by \decay{\jpsi}{\mup\mun} and nonresonant charm quark loops.

The local amplitudes of the \BdbToKstbMuMu decay are described within the framework of the Weak Effective Theory through the Hamiltonian~\cite{Buras:1998raa}
\begin{align}
    \mathcal{H} = \frac{-4 G_F}{\sqrt{2}} \lambda_t \sum_i \C{i} ^{(')} \Ope{i} ^{(')},
    \label{eq:WC}
\end{align}
where $G_F$ is the Fermi constant. The effective operators $\Ope{i} ^{(')}$ describe all possible interactions between the initial and final state. The relevant couplings for the \BdbToKstbMuMu decay are the Wilson coefficients \C7, \C9, and \C{10}, describing the strength and phase of the electromagnetic, vector, and axial-vector couplings, respectively, and their right-handed primed counterparts. Within the SM framework, all the Wilson coefficients have zero imaginary components.

The \BdbToKstbMuMu differential decay rate is defined in a five-dimensional phase space in terms of the three angular variables $(\cos\theta_\ell, \cos\theta_K, \phi)$, $q^2$, and the $\Km\pip$ mass squared $m_{K\pi}^2$. The angular conventions are the same as in Ref.~\cite{LHCb-PAPER-2013-019}. 
The \Kstarzb is reconstructed in the mass window \mbox{$796 < m_{K\pi} < 996\mevcc$}, and the final state receives contributions from P- and to a small extent S-wave amplitudes~\cite{LHCb-PAPER-2025-041}. 

Following the expressions given in Ref.~\cite{LHCb-PAPER-2024-011}, the \BdbToKstbMuMu decay amplitudes can be written as
\begin{equation}
    \begin{split}
        \displaystyle \mathcal{A}^{L,R}_{\lambda} =\mathcal{N}_{\lambda}(q^2) \lambda_t \bigg\{ \Big[ (\mathcal{C}_{9\,\lambda}^{\rm eff}(\qsq) \pm \mathcal{C}_{9}^{'})\mp (\mathcal{C}_{10} \pm \mathcal{C}_{10}^{'}) \Big]\mathcal{F}_{\lambda}(q^2)\\ + k_{\lambda}(q^2)\Big[(\mathcal{C}_{7}+\Delta\mathcal{C}_{7\,\lambda}\pm\mathcal{C}_{7}^{'})\Big]\mathcal{T}_{\lambda}(q^2)\bigg\},
    \end{split}\label{eq:amplitudes}
\end{equation}
where $\lambda$ denotes the polarisation of the $\Kstarzb$ meson and the factors $\mathcal{N}_{\lambda}(q^2)$, $k_{\lambda}(q^2)$, and the data-driven hadronic parameters $\Delta\mathcal{C}_{7\,\lambda}$ are defined in Ref.~\cite{LHCb-PAPER-2024-011}. The $\Bbar\to \Kstarzb$ form factors $\mathcal{F}_{\lambda}$, $\mathcal{T}_{\lambda}$ are taken from Ref.~\cite{Gubernari:2022hxn}. The sign in front of the right-handed Wilson coefficient depends on the polarisation of the \Kstarzb meson, whereas the amplitude chirality determines the overall sign in front of the axial-vector part of the amplitude. 
The analysis incorporates nonlocal amplitudes involving charm ($c\bar{c}$) and light-quark ($q\bar{q}$) loops by introducing an effective Wilson coefficient
\begin{equation}
    \begin{split}
        \mathcal{C}_{9\,\lambda}^{\rm eff} (\qsq) &= {\mathcal C_9} + \frac{\lambda_c}{\lambda_t} \left[ Y_{c\bar{c}} (q^2_0) + \frac{\left( \qsq - q^2_0 \right)}{\pi} \int^\infty_{4m_\mu^2} \frac{\rho_{c \bar{c}, \lambda} (s)}{(s - q^2_0) (s - \qsq - i \epsilon)} \deriv s \right] \\
        &+ \frac{\lambda_u}{\lambda_t}  \frac{\left( \qsq - q^2_0 \right)}{\pi} \int^\infty_{4m_\mu^2} \frac{\rho_{q \bar{q}, \lambda} (s)}{(s - q^2_0) (s - \qsq - i \epsilon)} \deriv s, 
    \end{split}
    \label{eq:dispersionRelation}
\end{equation}
that gives \C9 a $\qsq$ dependence through one-time subtracted dispersion relations for $c\bar{c}$ and $q\bar{q}$. The subtraction constant, $Y_{c\bar{c}}(q^{2}_{0})=\lambda_t/\lambda_c (0.174 + i0.006)$, at the point $q_{0}^{2}=-4.6\gevgevcccc$ is taken from the computations of Ref.~\cite{Asatrian:2019kbk}. The renormalisation scale is set to the \bquark-quark mass, $m_b=4.2\gevcc$ and  $m_{c}/m_{b}=0.29$. For $q\bar{q}$ contributions, the same subtraction point is used, and the subtraction constant is set to zero. This assumption has a negligible effect on the measurement, as reported in Ref.~\cite{LHCb-PAPER-2024-011}. Nonlocal contributions are added through the polarisation state $\lambda$-dependent spectral densities, denoted as $\rho_{c\bar{c},\,q\bar{q}}$, and include single-particle resonances corresponding to the $\rho(770)$, $\omega$, $\phi(1020)$, \jpsi, \psitwos, $\psi(3770)$, $\psi(4040)$ and $\psi(4160)$ mesons and two-particle open-charm intermediate states, parametrised following the recipe outlined in Refs.~\cite{LHCb-PAPER-2024-011,Cornella:2020aoq}. In particular, conservation of parity in the production of the two-particle intermediate states dictates that the two-particle system is generated in a P-wave configuration. Under this assumption, the three two-particle channels, $D^*\Dbar$, $D\Dbar$, and $D^*\Dbar^*$, can be approximated by a single effective $D^*\Dbar$ contribution, due to the similarity of the \qsq-dependent behaviour of the corresponding two-particle amplitudes. The sensitivity to \CP violation arises from the weak phase difference between the \C9 term and the term containing $\rho_{c \bar{c}}$ in Eq.~\ref{eq:dispersionRelation}.

As can be seen in Eqs.~\ref{eq:dispersionRelation} and~\ref{eq:amplitudes}, the phases of \C9 and \C{10} are measured relative to the SM weak phase difference between the local and nonlocal amplitudes. The combined fit to all CKM matrix elements~\cite{CKMfitter2015} gives this SM value as \mbox{$\arg(\lambda_t/\lambda_c) = 0.0189\pm0.0003$}, which is used in the results presented here.

The LHCb detector~\cite{LHCb-DP-2008-001,LHCb-DP-2014-002} is a single-arm forward spectrometer covering the pseudorapidity range $2<\eta<5$. Its tracking system comprises a silicon-strip vertex detector (VELO) surrounding the interaction region~\cite{LHCb-DP-2014-001}, a large-area silicon-strip detector upstream of a dipole magnet with a bending power of about $4\,\mathrm{T\,m}$, and three downstream tracking stations of silicon strips and straw drift tubes~\cite{LHCb-DP-2017-001}, providing precise momentum and vertex reconstruction. The magnetic field deflects oppositely charged particles in opposite directions and this can lead to detection asymmetries. Periodically reversing the magnetic field polarity throughout the data taking almost cancels the effect. 
Particle identification is achieved using ring-imaging Cherenkov detectors~\cite{LHCb-DP-2012-003} and a muon system of alternating iron absorbers and multiwire proportional chambers. A two-stage trigger, with a hardware stage selecting high-transverse-momentum muons and a software stage performing full event reconstruction, is used for efficient event selection. 

This is the first \CP violation measurement that exploits the angular distribution of \BdbToKstbMuMu candidates in the full \qsq spectrum, including regions dominated by charmonium intermediate states. It uses \proton\proton data collected by the \lhcb collaboration during 2011--2012 (Run 1) and 2016--2018 (Run 2), corresponding to an integrated luminosity of 8.4\invfb.

Simulated samples are used to determine the signal candidate selection requirements and to evaluate the efficiency. In the simulation, $pp$ collisions are generated using \pythia~\cite{Sjostrand:2007gs} with a specific \lhcb configuration~\cite{LHCb-PROC-2010-056}. Decays of unstable particles are described by \evtgen~\cite{Lange:2001uf}, in which final-state radiation is generated using \photos~\cite{davidson2015photos}. The interaction of the generated particles with the detector and its response are implemented using the \geant toolkit~\cite{Allison:2006ve, *Agostinelli:2002hh} as described in Ref.~\cite{LHCb-PROC-2011-006}. To ensure agreement between the simulation and the data, independent samples are used to calibrate the simulation and correct for potential discrepancies.

The selection of \BdbToKstbMuMu candidates follows Ref.~\cite{LHCb-PAPER-2024-011}, using the same selection requirements described therein. The data are selected within the range $0.1 \leq\qsq \leq 18.0\gevgevcccc$ to avoid the phase-space drop at the low edge of the $\qsq$ range and an increased \qsq dependence of the $m_{K\pi}$ lineshape at the high edge. 

After the initial selection, one of the main sources of background is due to random combinations of charged particles that do not originate from a single decay, referred to as combinatorial background. This is suppressed by applying a requirement on the output of a boosted decision tree~(BDT) classifier~\cite{Breiman,AdaBoost} trained for the $\qsq$ binned analysis in Ref.~\cite{LHCb-PAPER-2025-041}. This achieves a signal efficiency of approximately 87\% and 90\% in Run~1 and Run~2, respectively. The remaining combinatorial background is modelled and determined in the fit. 

Peaking backgrounds arise from misidentified decays mimicking the signal. The main ones are as follows: $\decay{\Bub}{\Km\mumu}$ combined with a random $\pip$, $\decay{\Bsb}{\phi(1020)}$ decays in which a charged kaon from the $\phi(1020)$ decay is misidentified as a pion; nonresonant $\decay{\Bsb}{\Kp\Km\mumu}$ decays, and $\decay{\Bdb}{\pip\pim\mumu}$ decays involving pion--kaon misidentification. Further contributions arise from $\decay{\Lb}{\proton \Km \mumu}$ and $\decay{\Lb}{\proton \pim \mumu}$ decays with proton misidentification, $\decay{\Bdb}{\jpsi\Kstarzb}$ and $\decay{\Bdb}{\psitwos\Kstarzb}$ decays with muon--hadron misidentification, and partially reconstructed $\decay{\Bub}{\Kstarm\mumu}$ decays.
These backgrounds are mitigated by applying mass vetoes under alternative mass hypotheses for the final-state particles, stringent particle-identification (PID) criteria, and multivariate selections based on BDT classifiers trained on simulated samples.

The kinematic phase space of the selected candidates is sculpted by the detector acceptance, the event reconstruction and the final-state selection. This effect is accounted for using an acceptance function derived from simulation, and modelled using Legendre polynomials in the phase space (\qsq, $\cos{\theta_\ell}, \cos{\theta_K}, \phi$).

The smearing of the true dimuon mass due to the experimental resolution is accounted for using resolution models, defined in three \qsq regions containing the $\phi(1020), \, \jpsi$, and \psitwos narrow resonances, $0.10 < \qsq < 3.24 \gevgevcccc$, $3.24 < \qsq < 11.56 \gevgevcccc$ and $11.56 < \qsq < 18.00 \gevgevcccc$.
Each model is described by the sum of a Gaussian and two Crystal Ball~(CB) functions~\cite{Skwarnicki:1986xj}, whose parameters are extracted from data in the $\qsq$ regions surrounding the \jpsi and \psitwos resonances and for the $\qsq$ region around the $\phi(1020)$ from simulation. The impact of resolution smearing on the angular observables is found to be negligible.

The measurement involves a sequence of unbinned maximum-likelihood fits performed to the mass and angular distributions of the \BdbToKstbMuMu decay. All fits are performed simultaneously to the \Bz and \Bzb samples. The procedure begins with a one-dimensional fit to the $m(\Kp\pim\mumu)$ mass distribution to determine the signal fraction,  using candidates in the range $5220 \leq m(\Kp\pim\mumu) \leq 5840 \mevcc$. The signal fraction is determined separately for \Bz and \Bzb candidates in five distinct \qsq intervals. These correspond to the three \qsq regions defined above, with the mid- and high-\qsq regions further subdivided in order to isolate the resonance-dominated contributions. The reconstructed signal mass distribution is modelled as the sum of two CB functions, with power-law tails on opposite sides of the peak. 

The signal region is defined as $5259.58 < m(\Km\pip\mumu) < 5299.58$ \mevcc and in this region the candidates are refit to impose a mass constraint of $m(\Kp\pim\mumu)$ to the known $B^0$ mass.
The combinatorial background distribution is determined from the upper mass sideband $m(\Kp\pim\mumu) > 5440 \mevcc$ and separately for each of the three \qsq ranges. The combinatorial background shape is modelled in the \ctl, \ctk, \phih, and \qsq dimensions, and each dimension is modelled independently. The sideband is divided into ten intervals, each of width 40\mevcc and with $m(\Kp\pim\mumu)$ constrained to the centre of the region. This is to mimic the mass constraint applied to signal events and, thus, ensure that peaking background components have the same shape in the signal region and the sidebands.

The final fit is performed to the angular distribution in the four-dimensional space (\ctl, \ctk, \phih, \qsq) in the signal region. In this step, the Wilson coefficients, including the magnitudes and phases of \C{9} and \C{10} , and the real parts of the opposite-chirality parameters \Cp{9} and \Cp{10} are determined. The magnitudes of \C{9} and \C{10} are the same for \Bzb and \Bz, while the corresponding phases $\DTV$ and $\DTAV$ are added to the \Bzb and subtracted from the \Bz. The phase of \C7 is assumed to be negligible and the magnitude of \C7 is fixed to its SM value, consistent with constraints from radiative $B$ decays~\cite{HFLAV23}. The phases of \Cp9 and \Cp{10} are also assumed to be negligible. While both the dominant P-wave and subdominant S-wave amplitudes for the $\Kp\pim$ system depend on the Wilson coefficients, their contributions to the S-wave are treated as nuisance parameters to avoid additional systematic effects from the poorly known form factors for the S-wave amplitudes.

\begin{figure}
    \centering
    \includegraphics[width=0.5\linewidth]{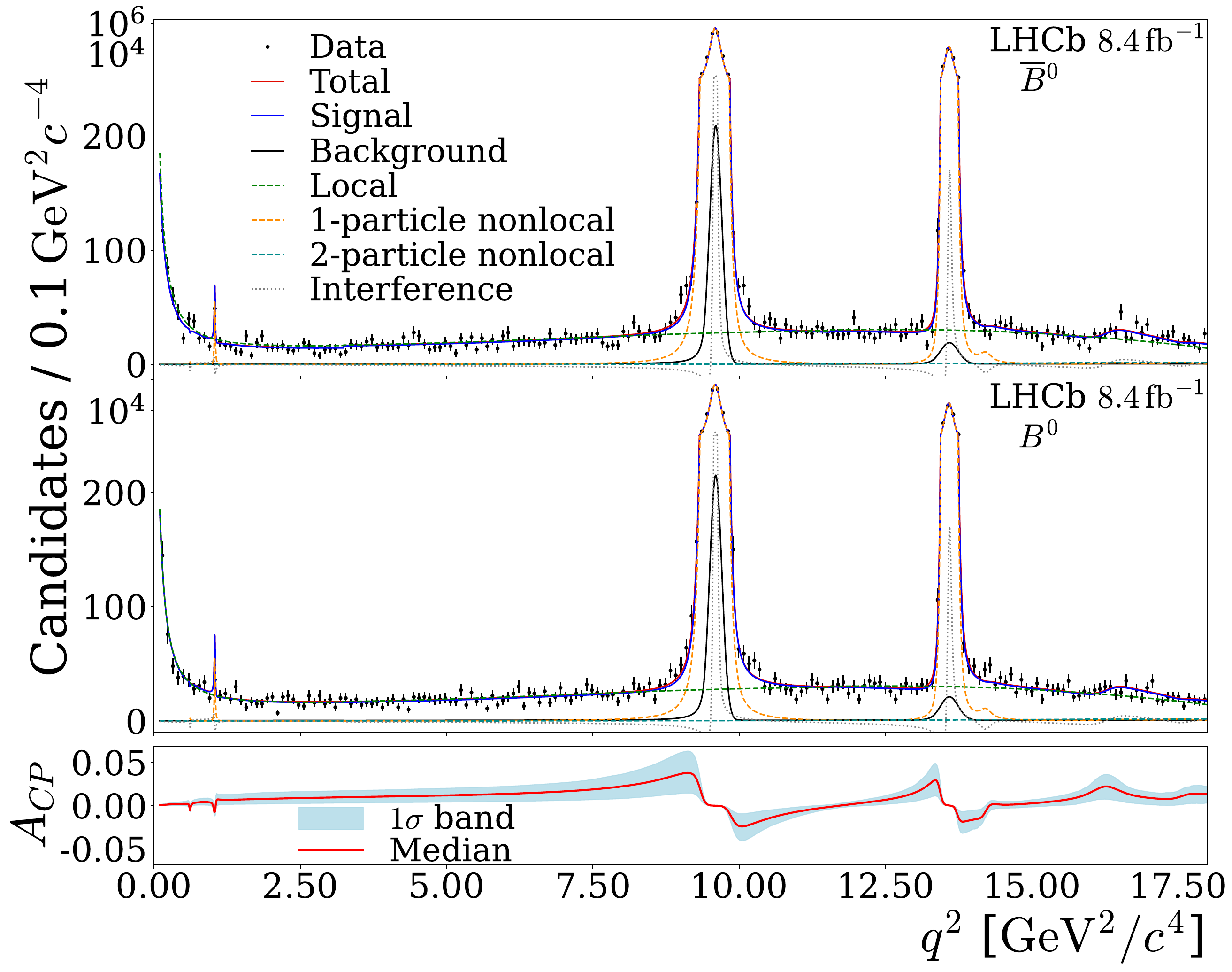}
    \caption{The \qsq distributions of the data and the corresponding fit, for the (top) \Bzb  and (middle) \Bz candidates. Note that the top of the $y$-axes scales are logarithmic to fit in the peaks of the charmonium resonances. The bottom panel displays the direct \CP asymmetry, $A_{\CP}=\frac{\Gamma(\Bzb)-\Gamma(\Bd)}{\Gamma(\Bzb)+\Gamma(\Bd)}$, as a function of \qsq, derived from the fit. The red line indicates the central value, and the blue band the $\pm1\sigma$ uncertainty.}
    \label{fig:q2spectrum}    
\end{figure}

The probability density function describing the signal decay rate is convolved with the $q^2$ resolution function and multiplied by an acceptance function to account for detector efficiency and response. The signal components for both \Bz and \Bzb are normalised using the measured branching fractions of the decay $\decay{\Bz}{\jpsi\Kstarz}$ followed by $\decay{\jpsi}{\mumu}$. This normalisation procedure renders the fit insensitive to potential production asymmetries between \Bz and \Bzb mesons. The final fit has 153 free parameters in total. This includes Wilson coefficients, parameters of nonlocal effects (one-, two-particle contributions and shifts to \C{7} ), local form factors, parameters related to the resolution in \qsq, parameters describing the combinatorial background, and parameters describing the signal fraction and $m(\Km\pip\mumu)$ distribution. The analysis framework follows the previous \CP-averaged analysis of the same dataset~\cite{LHCb-PAPER-2024-011}, where the detailed implementation of the common modelling ingredients is described.
Theoretical predictions~\cite{Gubernari:2022hxn} are used as Gaussian constraints on the local form factors in the fit. The signal fraction and the combinatorial background shape are constrained using the results from the previous fits to the $m(\Km\pip\mumu)$ spectra and the angular distribution in the background sidebands. The fit projection on \qsq and the resulting direct \CP violation are shown in Fig.~\ref{fig:q2spectrum}.

Results are consistent when splitting the dataset by year and magnet polarity and performing fits on those individually. Additionally, the fit is applied to five randomly selected subsets of the data. To verify the absence of spurious \CP violation, two tests are performed in which the $\Bz/\Bzb$ tag is artificially changed and the nominal fit repeated. In the first the tag is random, and in the second it is based on magnet polarity. As expected, the obtained result is consistent with no \CP violation in both cases.

Differences in the detection efficiency of the \Bzb and \Bz final states arise as the detector itself is not matter-antimatter symmetric. The largest difference is in the interactions of positive and negative kaons. This detection asymmetry may contribute to the measured \CP violation. The associated systematic uncertainty is studied using pseudoexperiments. First, the nominal acceptance function is determined using simulation without distinguishing \Bzb and \Bz candidates. 
To evaluate the systematic uncertainty arising from the acceptance difference, separate acceptance functions are calculated for \Bzb and \Bz mesons. These functions are derived with the kaon PID distributions corrected using data from decays where the kaon can be identified directly from kinematics. The systematic uncertainty of the acceptance difference between \Bzb and \Bz is derived by using the two acceptance functions for generating pseudoexperiments but using the nominal one for the fit. The resulting systematic uncertainty is 22\%~(13\%) relative to the statistical uncertainty for $\DTV$~($\DTAV$\kern-0.5pt). In a similar way, any possible difference in $\qsq$ resolution between the \Bzb and \Bz events is considered.

In addition to the systematic uncertainties directly relevant to the \CP violation measurement, all sources considered in the \CP-averaged analysis~\cite{LHCb-PAPER-2024-011} are reevaluated in this study. For the magnitude of the Wilson coefficients, the uncertainty in the branching fraction of the $\decay{\Bzb}{\jpsi\Kstarzb}$ decay dominates, while for the \CP violating parameters, it is the uncertainty in the background from $B\to T_{c\overline{c}1}(\to\psi\pi)K$ decays. 

\begin{figure}[t]
    \centering
    \includegraphics[width=0.45\linewidth]{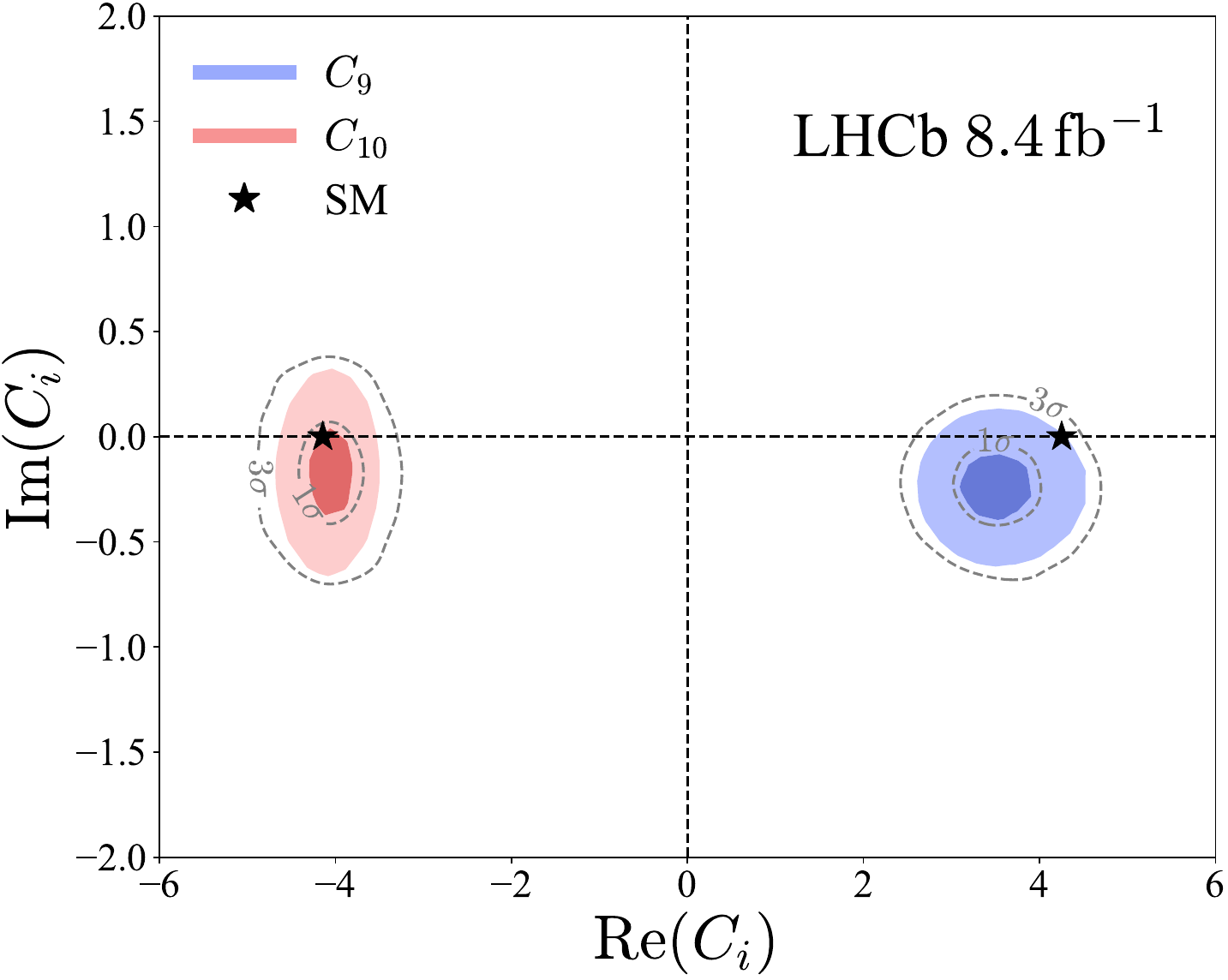}
    \caption{Two-dimensional likelihood profiles of the Wilson coefficients \C{9} (blue) and \C{10} (red) in the complex plane. The profiles are shown using the statistical only (filled regions) and total statistical and systematic uncertainties (dotted lines). The one- (68\% CL) and three-sigma (99.7\% CL) regions are highlighted. The stars represent the SM predictions.}
    \label{fig:LHprofile}
\end{figure}

\begin{table}
\centering
\caption{Results of the fit for the Wilson coefficients. The top part of the table shows the magnitude and phase; the bottom one shows the real and imaginary components. In each row, the first uncertainty is statistical and the second is systematic.}
\begin{tabular}{l  r }
    \toprule
    \multicolumn{2}{c}{Wilson coefficient results} \\
    \midrule
    $\abs{\mathcal{C}_9}$ & $3.50 \pm 0.30\phantom{0} \pm 0.15\phantom{0}$ \\
    $\abs{\mathcal{C}_{10}}$ & $4.04 \pm 0.18\phantom{0} \pm 0.15\phantom{0}$ \\
    $\DTV$ & $-0.067 \pm 0.032 \pm 0.011$ \\
    $\DTAV$ & $0.043 \pm 0.035 \pm 0.010$ \\
    \midrule
    $\Real(\mathcal{C}_9)$ & $3.50 \pm 0.30\phantom{0}  \pm 0.15\phantom{0}$ \\
    $\Imag(\mathcal{C}_9)$ & $-0.23 \pm 0.11\phantom{0}  \pm 0.04\phantom{0}$ \\
    $\Real(\mathcal{C}_{10})$ & $-4.04 \pm 0.18\phantom{0}  \pm 0.15\phantom{0}$ \\
    $\Imag(\mathcal{C}_{10})$ & $-0.17 \pm 0.14\phantom{0}  \pm 0.04\phantom{0}$ \\
    $\Real(\mathcal{C}_9^\prime)$ & $\phantom{-}0.39 \pm 0.46\phantom{0} \pm 0.21\phantom{0}$ \\
    $\Real(\mathcal{C}_{10}^\prime)$ & $-0.12 \pm 0.21\phantom{0} \pm 0.06\phantom{0}$ \\
    \bottomrule
\end{tabular}
\label{tab:WCResults}
\end{table}

A summary of the determined values of the Wilson coefficients together with the statistical and systematic uncertainties is given in Table~\ref{tab:WCResults}. Correlation contours in the complex plane of \C{9} and \C{10} are shown in Fig.~\ref{fig:LHprofile}. These are generated from a profile likelihood scan in the 2D plane and, thus, reflect the simultaneous fit to all the other model parameters. In particular, the other Wilson coefficients of the model are not assumed to take on the SM values during the profiling procedure.

The analysis reports no significant evidence for direct \CP violation in the \mbox{\BdbToKstbMuMu} decay. Under the assumption that the SM is the only contribution to \CP violation in the decay, the weak phase for \C9 and \C{10} can be averaged to give \mbox{$-0.012\pm0.025 \stat \pm 0.008 \syst$}. This can be compared to the measurement from \Bs oscillations in the decay $\decay{\Bsb}{\jpsi\phi}$~\cite{HFLAV23}, which gives $-0.007 \pm 0.007$~\cite{HFLAV23} after offsetting it with the indirect measurement of $\arg(\lambda_t/\lambda_c)$ to take into account the definition of the Wilson coefficients in Eq.~\ref{eq:WC}. 

For the parameters that are shared, the result is consistent with the \CP-conserving unbinned analysis of the same decay, and the measured angular observables are in good agreement with the recent binned \BdbToKstbMuMu study~\cite{LHCb-PAPER-2025-041}. The favouring of $\Real(\mathcal{C}_9)$ with a value that is around $-0.8$ smaller than the SM expectation is consistent with other analyses but has a lower significance of $2.1\sigma$ because the effect of nonlocal contributions is taken from the fit rather than estimated from theory and because all the other Wilson coefficients are not assumed to take on the SM values. The other Wilson coefficients determined are all compatible with the SM expectation within $1\sigma$.

In conclusion, a search for \CP violation has been performed in the unbinned $q^2$ range \mbox{$0.1 < \qsq < 18.0~\gevgevcccc$} of the \BdbToKstbMuMu decay using the combined Run~1 and Run~2 LHCb data. This analysis provides the first direct measurements of the phases of the $\bsll$ Wilson coefficients \C9 and \C{10} . Compared to determinations of the imaginary components of the Wilson coefficients from prior analyses conducted in bins of \qsq and not including the regions very close to the charmonium resonances, the resolution is improved by more than an order of magnitude~\cite{Fleischer:2025ucq}. The same is true for the determination of the weak phase of the Wilson coefficients~\cite{LHCb-PAPER-2025-062}. This Letter presents the first analysis that is sensitive to \CP violation in flavour-changing neutral-current decays at a level similar to the expectation from the Standard Model. It also explores a set of \CP-violating couplings that \Bs oscillations are not sensitive to.

%% file: acknowledgements.tex
\section*{Acknowledgements}
%
%
\noindent We express our gratitude to our colleagues in the CERN
accelerator departments for the excellent performance of the LHC. We
thank the technical and administrative staff at the LHCb
institutes.
We acknowledge support from CERN and from the national agencies:
ARC (Australia);
CAPES, CNPq, FAPERJ and FINEP (Brazil); 
MOST and NSFC (China); 
CNRS/IN2P3 and CEA (France);  
BMFTR, DFG and MPG (Germany);
INFN (Italy); 
NWO (Netherlands); 
MNiSW and NCN (Poland); 
MEC/IFA (Romania); 
MICIU and AEI (Spain);
SNSF and SER (Switzerland); 
NASU (Ukraine); 
STFC (United Kingdom); 
DOE NP and NSF (USA).
We acknowledge the computing resources that are provided by ARDC (Australia), 
CBPF (Brazil),
CERN, 
IHEP and LZU (China),
IN2P3 (France), 
KIT and DESY (Germany), 
INFN (Italy), 
SURF (Netherlands),
Polish WLCG (Poland),
IFIN-HH (Romania), 
PIC (Spain), CSCS (Switzerland), 
GridPP (United Kingdom),
and NSF (USA).  
We are indebted to the communities behind the multiple open-source
software packages on which we depend.
Individual groups or members have received support from
RTP (Australia), 
FWO Odysseus grant G0ASD25N (Belgium), 
Key Research Program of Frontier Sciences of CAS, CAS PIFI, CAS CCEPP (China); 
Minciencias (Colombia);
EPLANET, Marie Sk\l{}odowska-Curie Actions, ERC and NextGenerationEU (European Union);
A*MIDEX, ANR, IPhU and Labex P2IO, and R\'{e}gion Auvergne-Rh\^{o}ne-Alpes (France);
Alexander-von-Humboldt Foundation (Germany);
ICSC (Italy); 
Severo Ochoa and Mar\'ia de Maeztu Units of Excellence, GVA, XuntaGal, GENCAT, InTalent-Inditex and Prog.~Atracci\'on Talento CM (Spain);
the Leverhulme Trust, the Royal Society and UKRI (United Kingdom).

%% file: Authorship_LHCb-PAPER-2026-007.tex
\centerline
{\large\bf LHCb collaboration}
\begin
{flushleft}
\small
R.~Aaij$^{38}$\lhcborcid{0000-0003-0533-1952},
M.~Abdelfatah$^{69}$,
A.S.W.~Abdelmotteleb$^{57}$\lhcborcid{0000-0001-7905-0542},
C.~Abellan~Beteta$^{51}$\lhcborcid{0009-0009-0869-6798},
F.~Abudin\'en$^{59}$\lhcborcid{0000-0002-6737-3528},
T.~Ackernley$^{61}$\lhcborcid{0000-0002-5951-3498},
A.A.~Adefisoye$^{69}$\lhcborcid{0000-0003-2448-1550},
B.~Adeva$^{47}$\lhcborcid{0000-0001-9756-3712},
M.~Adinolfi$^{55}$\lhcborcid{0000-0002-1326-1264},
P.~Adlarson$^{87,42}$\lhcborcid{0000-0001-6280-3851}
% ...

{\footnotesize \it

% ...
$^{38}$Nikhef National Institute for Subatomic Physics, Amsterdam, Netherlands\\
% ...
$^{42}$National Center for Nuclear Research (NCBJ), Warsaw, Poland\\
% ...
$^{47}$Instituto Galego de F{\'\i}sica de Altas Enerx{\'\i}as (IGFAE), Universidade de Santiago de Compostela, Santiago de Compostela, Spain\\
% ...
$^{51}$Physik-Institut, Universit{\"a}t Z{\"u}rich, Z{\"u}rich, Switzerland\\
% ...
$^{55}$H.H. Wills Physics Laboratory, University of Bristol, Bristol, United Kingdom\\
% ...
$^{57}$Department of Physics, University of Warwick, Coventry, United Kingdom\\
% ...
$^{59}$School of Physics and Astronomy, University of Edinburgh, Edinburgh, United Kingdom\\
$^{60}$School of Physics and Astronomy, University of Glasgow, Glasgow, United Kingdom\\
$^{61}$Oliver Lodge Laboratory, University of Liverpool, Liverpool, United Kingdom\\
% ...
$^{69}$Syracuse University, Syracuse, NY, United States\\
% ...
$^{87}$Department of Physics and Astronomy, Uppsala University, Uppsala, Sweden, associated to $^{60}$\\
% ...
}
\end{flushleft}